\documentclass[sigconf, authorversion]{acmart}
\usepackage{enumitem, listings}
\usepackage{xcolor} 
\usepackage{framed}            

\def\BibTeX{{\rm B\kern-.05em{\sc i\kern-.025em b}\kern-.08emT\kern-.1667em\lower.7ex\hbox{E}\kern-.125emX}}

\copyrightyear{2024}
\acmYear{2024}
\setcopyright{acmlicensed}
\acmConference[ITiCSE 2024]{Proceedings of the 2024 Innovation and Technology in Computer Science Education V. 1}{July 8--10, 2024}{Milan, Italy}
\acmBooktitle{Proceedings of the 2024 Innovation and Technology in Computer Science Education V. 1 (ITiCSE 2024), July 8--10, 2024, Milan, Italy}
\acmDOI{10.1145/3649217.3653568}
\acmISBN{979-8-4007-0600-4/24/07}

\graphicspath{{Graphs/}}
\begin{document}

%

\title[Automating Personalized Parsons Problems]{Automating Personalized Parsons Problems with Customized Contexts and Concepts}

\author{Andre del Carpio Gutierrez}
\orcid{0009-0004-0208-4567}
\affiliation{
  \institution{University of Auckland}
  \city{Auckland}
  \country{New Zealand}
}
\email{adel868@aucklanduni.ac.nz}

\author{Paul Denny}
\orcid{0000-0002-5150-9806}
\affiliation{
  \institution{University of Auckland}
  \city{Auckland}
  \country{New Zealand}
}
\email{paul@cs.auckland.ac.nz}

\author{Andrew Luxton-Reilly}
\orcid{0000-0001-8269-2909}
\affiliation{
  \institution{University of Auckland}
  \city{Auckland}
  \country{New Zealand}
}
\email{andrew@cs.auckland.ac.nz}

\renewcommand{\shortauthors}{Andre del Carpio Gutierrez et al.}


%



%

%

\begin{abstract}
Parsons problems provide useful scaffolding for introductory programming students learning to write code. However, generating large numbers of high-quality Parsons problems that appeal to the diverse range of interests in a typical introductory course is a significant challenge for educators.  Large language models (LLMs) may offer a solution, by allowing students to produce on-demand Parsons problems for topics covering the breadth of the introductory programming curriculum, and targeting thematic contexts that align with their personal interests.  In this paper, we introduce PuzzleMakerPy, an educational tool that uses an LLM to generate unlimited contextualized drag-and-drop programming exercises in the form of Parsons Problems, which introductory programmers can use as a supplemental learning resource. We evaluated PuzzleMakerPy by deploying it in a large introductory programming course, and found that the ability to personalize the contextual framing used in problem descriptions was highly engaging for students, and being able to customize the programming topics was reported as being useful for their learning. 

\end{abstract}

%
%

%

%

%
\maketitle

\section{Introduction}
The primary goal for students in introductory programming courses has traditionally been to learn how to write code. Beginners often find it challenging and time-consuming to write code from scratch~\cite{qian_students_2018, benda_when_2012}, and thus effective activities to scaffold code writing are valuable.  Parsons problems have become one of the most widely studied and popular scaffolding tasks in introductory courses \cite{ericson_parsons_2022}.
They are programming puzzles where students rearrange incorrectly ordered code fragments into a correct working program ~\cite{parsons_parsons_2006}. Although Parsons problems have several benefits for students~\cite{ericson_parsons_2022}, it is time-consuming for educators to create large numbers of these educational resources ~\cite{mirhosseini_what_2023}. In addition, we believe that the learning experience is likely to be improved when students have the autonomy to personalize their learning resources by selecting problem contexts that align with their interests
\cite{leinonen2021exploring}.  Large language models (LLMs) may help with automating the creation of these exercises, as they have been shown capable of producing thematic exercises (i.e., generating a context for a given problem) and their associated code solutions ~\cite{sarsa_automatic_2022}. 

In this research, we describe PuzzleMakerPy, an educational tool that uses an LLM (OpenAI's GPT-3.5) to automatically create an unlimited number of personalized Parsons problems in which the context and programming concepts can be tailored to individual students. 
We deployed 
PuzzleMakerPy in a large introductory programming course, and evaluated its use guided by the following research questions:
\begin{itemize}[topsep=2pt]
    \item \textbf{RQ1:} \emph{How do introductory programming students interact with the PuzzleMakerPy educational tool?}
    \item \textbf{RQ2:} \emph{What attitudes emerge from their engagement with the tool?}
\end{itemize}



We found that PuzzleMakerPy was well-received by students, who reported the ability to personalize Parsons problems via contexts and programming concepts enjoyable and useful for their learning. Furthermore, we found that students used a wide range of the provided contexts available to them in the tool, and would often prescribe their own contexts. We also found that students used all provided programming concepts but never entered their own, indicating that a reasonably sized list that covers typical introductory programming curricula may suffice for their learning.


\section{Related Work}
Extensive prior work exists on both the use and integration of LLMs in programming education and on Parsons problems.  

 \begin{figure*}[t!]
    \centering
    \includegraphics[width=13.5cm]{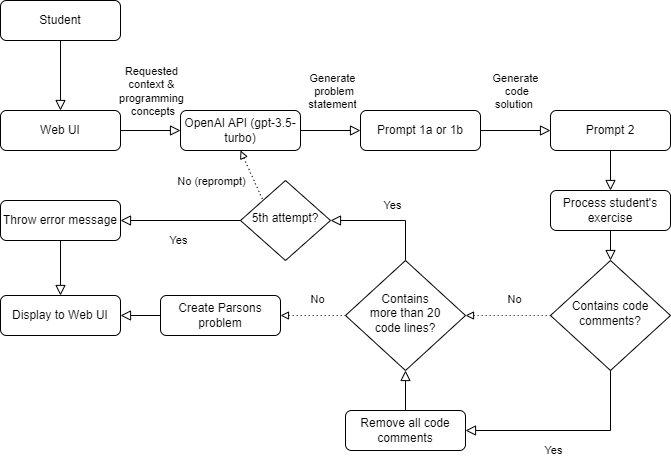}
    \caption{A diagram illustrating the architecture and information flow of PuzzleMakerPy.}
    \label{fig:tool_diagram}
\end{figure*}

\subsection{Large Language Models in Programming Education}

The integration of large language models (LLMs) into programming education has recently gained significant attention \cite{denny2023computing, prather2023robots}.  A growing body of literature now documents the capabilities of LLMs for solving a range of tasks relevant in computing courses, including generating solutions to programming problems \cite{finnie2022robots,denny2023conversing,hou2024more,savelka2023can}, test suite creation \cite{chen2022codet}, detecting bugs \cite{macneil2024decoding}, and explaining both code \cite{sarsa_automatic_2022, leinonen_comparing_2023} and error messages \cite{leinonen_using_2023, phung_generating_2023}.  LLM-powered tools are also entering computing classrooms as digital teaching assistants \cite{liffiton_codehelp_2023}, programming support tools \cite{kazemitabaar2023studying}, and tools to support the development of skills in prompt creation and engineering \cite{denny2024explaining, denny2024promptproblems}. Researchers have also explored the application of LLMs to generate learning resources for students.  Sarsa et al. explored the use of a pioneering LLM, Codex, for generating novel and sensible programming exercises that could be customised according to specified programming concepts and contextual themes \cite{sarsa_automatic_2022}.  Although much of the content generated using their few-shot approach was accurate, they suggested there was a need for oversight from instructors to ensure content quality.  Similar work on resource generation for programming courses has focused on multiple-choice questions \cite{doughty2024comparative} and worked examples \cite{jury2024evaluating}.

\subsection{Parsons Problems}
Numerous variations of Parsons problems have been explored, including puzzles that are adaptive ~\cite{ericson_evaluating_2018, hou_using_2022}, Parsons problems with or without distractors ~\cite{harms_distractors_2016}, single line puzzles related to regular expressions or SQL (micro) ~\cite{prather2023robots}, optional code lines ~\cite{ericson_solving_2017}, and faded Parsons problems (with incomplete code fragments that need to be filled) ~\cite{weinman_improving_2021, fromont2023exploring}. Parsons problems are known to provide many learning benefits for students such as improved learning engagement ~\cite{parsons_parsons_2006}, the ability to learn from immediate feedback ~\cite{ericson_parsons_2022, ericson_analysis_2015}, the potential to reduce cognitive load ~\cite{zhi_evaluating_2019}, and to result in similar learning gains when compared to writing code from scratch ~\cite{parsons_parsons_2006, zhi_evaluating_2019, ericson_solving_2017}.

In the current paper, we explore the automatic generation of Parsons problems.  Most similar to this work is the research of Hou et al., who present the CodeTailor tool, a system that uses LLMs to create personalized Parsons puzzles \cite{hou2024codetailor}.  In a study with 18 novice programmers, they found that most students preferred to use CodeTailor over receiving AI-generated solutions to support their learning. 
We previously evaluated the quality of LLM-generated  programming exercises spanning a broad range of contexts and programming concepts \cite{gutierrez2024evaluating}. The generated exercises included high-quality problem statements and code solutions, providing the necessary components to create Parsons problems for introductory programmers.

\section{Tool Design and Implementation}
The architecture (see Figure \ref{fig:tool_diagram}), design choices, technologies, and prompts used for PuzzleMakerPy are detailed in this section.

\begin{figure}
    \centering
    \includegraphics[width=8.5cm]{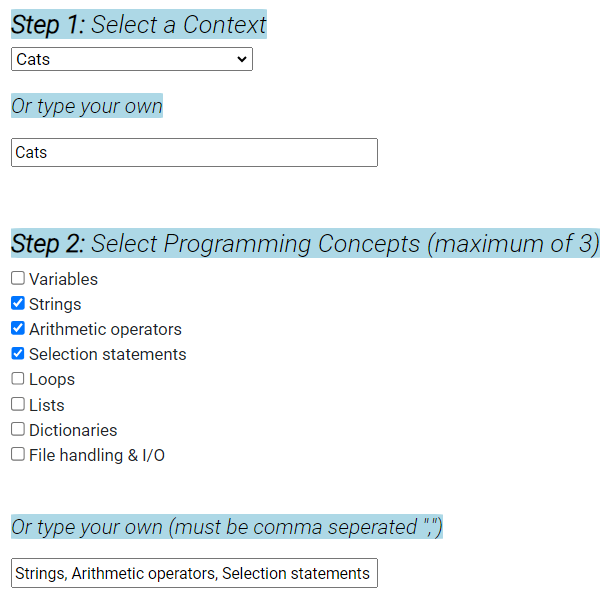}
    \caption{An image illustrating how a student would interact with PuzzleMakerPy when selecting their context and programming concepts to create a programming exercise.}
    \label{fig:tool_usage}
\end{figure}

\begin{figure}
    \centering
    \includegraphics[width=8.5cm]{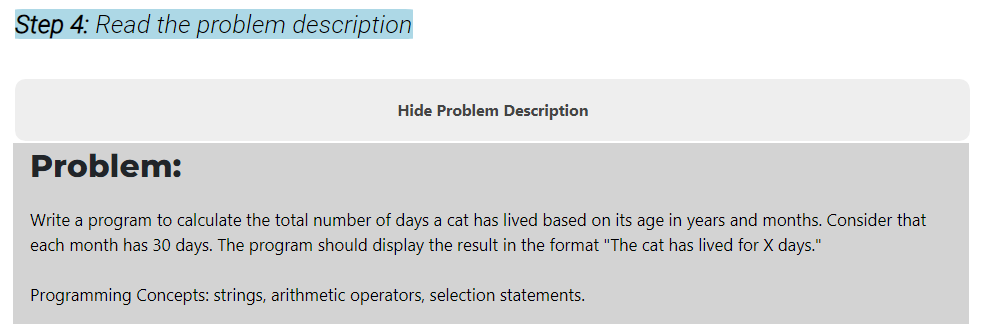}
    \caption{An image illustrating PuzzleMakerPy's interface for generated problem statements.}
    \label{fig:problem_statement}
\end{figure}

\begin{figure}
    \centering
    \includegraphics[width=8.5cm]{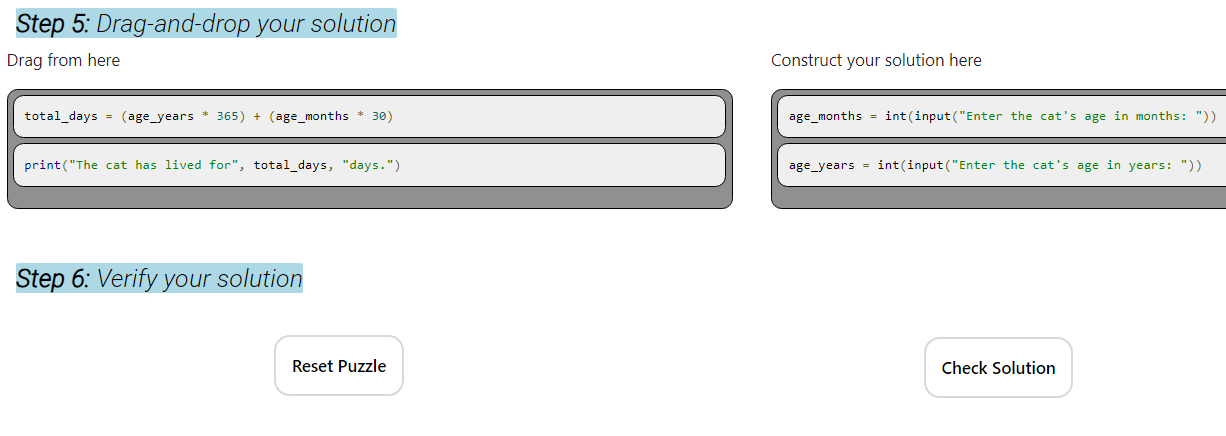}
    \caption{An image illustrating PuzzleMakerPy's drag-and-drop interface for code solutions.}
    \label{fig:solution_area}
\end{figure}

\begin{figure}
    \centering
    \includegraphics[width=8.5cm]{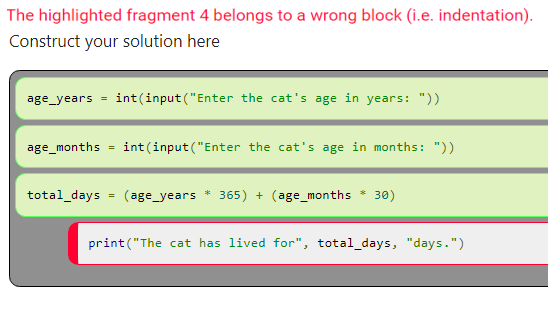}
    \caption{An image illustrating how PuzzleMakerPy provides feedback for a solution containing incorrect indentation.}
    \label{fig:tool_feedback}
\end{figure}

When students use PuzzleMakerPy, they are asked to personalize their problem by selecting a context and a set of programming concepts (see Figure \ref{fig:tool_usage}).  After generating the problem, students are provided with a problem description (see Figure \ref{fig:problem_statement}) and can use a typical drag-and-drop interface to develop their solutions (see Figure \ref{fig:solution_area}). Feedback on the submitted code solutions is also provided, as shown in Figure \ref{fig:tool_feedback}.

\subsection{Filtering Automated Content}
Figure \ref{fig:tool_diagram} provides an overview of the architecture of PuzzleMakerPy, which is designed to reliably create high-quality Parsons problems. The first design decision was limiting the number of code fragments in generated solutions through our zero-shot prompts (see Figure \ref{fig:tool_prompts}) and by reprompting only if solutions contained more than 20 lines of code. Reprompting for an exercise that fails to meet the conditions is performed five times consecutively before the student is asked to retry. This limit ensures that students are not stuck waiting, and reduces API costs if the LLM repeatedly fails to create the exercise. Another design choice included removing all code comments from exercises, as Parsons problems blocks typically do not contain comments. Finally, to reduce the likelihood of producing undesirable responses, our zero-shot prompts excluded various programming concepts in generated solutions, as these did not align with the learning objectives specific to the course that the tool was deployed in, and it may have confused students. We also included prompts to ensure that problem statements did not  contain extra unnecessary text.

\begin{figure}
\begin{framed}
\begin{minipage}{\columnwidth}
\footnotesize 
\textbf{Prompt 1a (problem description when a student requests for a context):} \\

Generate a problem in Python that only includes a problem statement with a context of ‘[\emph{context requested by a student goes here}],’ and the programming concepts should
be ‘[\emph{programming concepts requested by a student goes here}]’.

The solution associated with the problem should not exceed ten lines of code and should be limited to only a single function at most.

The problem cannot use \texttt{while(true)}, or any \texttt{break} statements or exceptions (\texttt{try, catch} blocks).

Do not include information on the constraints I have defined in the problem statement.

Only provide the problem statement as the output. \\

\textbf{Prompt 1b (problem description when a student requests for no context):} \\

Generate a problem in Python that only includes a problem statement with the following programming concepts: ‘[\emph{programming concepts requested by a student goes here}].’ Do not apply a context to the exercise.

The solution associated with the problem should not exceed ten lines of code and should be limited to only a single function at most.

The problem cannot use \texttt{while(true)}, or any \texttt{break} statements or exceptions (\texttt{try, catch} blocks).

Do not include information on the constraints I have defined in the problem statement.

Only provide the problem statement as the output. \\

\textbf{Prompt 2 (code solution):} \\

[\emph{problem description outputted from either prompt 1a or 1b}] \\

Only provide code solutions based on the problem description above.

Do not explain the solution or add any extra detail to the output; only provide the code solution.

\end{minipage}
\end{framed}
\caption{All three zero-shot prompts used for generating complete programming exercises within PuzzleMakerPy. Prompt 1a or Prompt 1b is used to generate the problem description, while Prompt 2 is solely used to generate code solutions.}
\label{fig:tool_prompts}
\end{figure}

\subsection{Technologies}
PuzzleMakerPy was developed using Flask (a micro web framework written in Python) and standard web technologies such as HTML, CSS, JavaScript, and jQuery. An external library that contributed to the creation of this educational tool is the open source \emph{js-parsons} library\footnote{\url{https://github.com/js-parsons/js-parsons}}.

The LLM used for prompting all programming exercises in PuzzleMakerPy was \emph{gpt-3.5-turbo}. Multiple zero-shot prompts were employed for exercise generation. The first  prompt focused on generating the problem description based on the context or thematic keyword provided by a student, denoted as either Prompt \emph{1a} or Prompt \emph{1b}. Prompt \emph{1a} handled requests for exercises that required a context, while Prompt \emph{1b} was used for exercises without a context. The second prompt (Prompt 2) handled the generation of code solutions, and this prompt was supplemented with the problem description obtained from either Prompt \emph{1a} or Prompt \emph{1b}. To observe the three zero-shot prompts for generating programming exercises requested by students, refer to Figure \ref{fig:tool_prompts}.



\subsection{Contexts and Programming Concepts}
Students were allowed to select contexts and programming concepts for their generated programs (see Figure \ref{fig:tool_usage}). To cover a wide range of student interests, 20 contexts (see Table \ref{table:tool_contexts}) were integrated into PuzzleMakerPy through a drop-down menu. In addition, students could choose to include either no context, type a specific context, or (by selecting a `Surprise me' option) use a random context selected from the topics of the Dewey Decimal Classification System (approximately 1000 topics). Introducing this level of flexibility is likely to align with students' interests, allowing them to discover enjoyable and engaging programming exercises.

Similarly, for programming concepts, students were given the choice to select a maximum of 3 concepts for generated solutions using checkboxes, as shown in Figure \ref{fig:tool_usage}. We believe that being able to combine multiple concepts for generated solutions would lead to more useful learning resources for students. For the programming concepts used (see Table \ref{table:tool_revisedconcepts}), students could also type their own concepts to be included before an exercise is created. However, unlike requesting exercises without context, students were not allowed to create exercises without programming concepts.

\begin{table}[hbt!]
\centering
\caption{List of contexts and programming concepts used in PuzzleMakerPy} 
\label{table:tool_contexts} \label{table:tool_revisedconcepts} 
\footnotesize
\begin{tabular} { p{8.0cm}}
 \toprule
  \multicolumn{1}{c}{\textbf{Contexts: 20}} \\
 \midrule
 Amusement Park; Animals; Aquarium; Basketball; Cooking; Film; Fishing; Gardening; Mental Health; Modern Gaming; Music; Olympics; Pets; Relationships; Restaurant; Rugby; Social Media; Sports; Streaming Services; Virtual Reality\\
  \midrule
\multicolumn{1}{c}{\textbf{Programming Concepts: 8}} \\
    \midrule
Arithmetic operators; Dictionaries; File handling \& I/O; Lists; Loops; Selection statements (if/else, etc.); Strings; Variables \\
\bottomrule
\end{tabular}
\end{table}


\section{Evaluation}
PuzzleMakerPy was deployed approximately halfway through an introductory programming course (CS1) with 423 students. Using the tool was optional for students, but they were encouraged to use it to revise their code-writing skills before a mid-semester test. Students in the course had not previously been taught using Parsons problems.

As use of the tool was anonymous, each problem request was logged independently, and so we can only report the number of attempts rather than the number of students using the tool. We logged elements of student behavior and asked them to complete an optional questionnaire.


\subsection{Student Requested Keywords}
Students had various options for selecting contexts and programming concepts before generating exercises, either from predefined lists (see Table \ref{table:tool_contexts}) or by typing their own.   We collected logs of use from PuzzleMakerPy and report the top 20 student-requested contexts in Table \ref{table:tool_top_contexts}. Additionally, the frequency of student-requested programming concepts is shown in Table \ref{table:concepts_requested}.

\begin{table}[htb]
\caption{Top 20 student-requested contexts in generated questions (out of 236). \emph{`None'} refers to no-context, \emph{`Surprise Me'} is a random context extracted from the Dewey Decimal System, and \emph{`Custom'} is a context typed by a student.}
\label{table:tool_top_contexts} 
\footnotesize
\begin{tabular} { p{2.5cm} p{1.0cm}}
 \toprule
 \textbf{Context Requested} & \textbf{Number of questions} \\
 \midrule
 Animals & 33 \\
 Music & 22 \\
 \emph{Custom} & 20 \\
 Basketball & 17 \\
 Cats & 17 \\
 Amusement Park & 14 \\
 Mental Health & 11 \\
 Modern Gaming & 9 \\
 Sports & 9 \\
 Film & 8 \\
 Fishing & 8 \\
 \emph{None} & 8 \\
 Relationships & 8 \\
 Cooking & 5 \\
 Olympics & 5 \\
 Pets & 5 \\
 \emph{Surprise Me} & 5 \\
 Gardening & 4 \\ 
 Restaurant & 4 \\
 Social Media & 3 \\
 \bottomrule
\end{tabular}
\end{table}

\begin{table}[htb]
\caption{The frequency of student-requested programming concepts in generated questions (out of 236)} 
\label{table:concepts_requested} 
\footnotesize
\begin{tabular} { p{2.5cm} p{1.5cm}}
 \toprule
 \textbf{Programming Concept Requested} & \textbf{Number of questions} \\
 \midrule
  Loops & 110 \\
 Variables & 106 \\
 Strings & 78 \\
 Lists & 77 \\
 Dictionaries & 73 \\
 Arithmetic Operators & 32 \\
 File Handling \& I/O & 30 \\
 Selection Statements & 17 \\
 \bottomrule
\end{tabular}
\end{table}

For requested contexts, \emph{`animals'} were the most popular (33 questions), 
and relatively few questions were generated with \emph{`none'} (8 questions), and even fewer using the \emph{`surprise me'} (5 questions) feature.  Many questions were also generated using \emph{`custom'} contexts (20 questions) which were entered by students, indicating there is value in including this option despite the extensive list of contexts provided.

For programming concepts requested by students, \emph{`loops'} were the most prevalent (110 questions), with \emph{`variables'} a close second (106 questions). Variables may appear high on the list as they are commonly used in programs and are a generally well-understood concept. Students used all of the programming concepts offered, although the distribution was not even, but surprisingly no student entered their own programming concepts.  This suggests that using a small list focused on the key concepts relevant to the course is likely sufficient. 

In summary, student behavior shows a wide range of contexts being used, with few opting not to use contextualization. Similarly, students used all of the predefined programming concepts when generating questions, although no questions were generated with student-entered topics.

\subsection{Student satisfaction}
After completing exercises using the tool, students were invited to respond to a short questionnaire.  Four Likert-scale questions asked students about their preferences, followed by four open-ended questions that asked students to reflect on their experiences.  The questions are shown in Tables \ref{table:tool_likert_survey} and \ref{table:tool_openended_survey}.

\begin{table}[htb]
\caption{Likert scale survey questions and results (\emph{n} = 18) Abbreviations: SD = Strongly Disagree, D = Disagree, N = Neutral, A = Agree, SA = Strongly Agree} 
\label{table:tool_likert_survey} 
\footnotesize
\begin{tabular} { p{3.2cm} r r r r r }
 \toprule
\textbf{Question} & \textbf{SD} & \textbf{D} & \textbf{N} & \textbf{A} & \textbf{SA} \\
 \midrule
Q1. Customizing the context of a programming question was interesting/enjoyable.
& 0\% & 0\% & 0\% & 50\% & 50\%  \\
 \midrule 
Q2. Customizing programming concepts is valuable for improving my own programming skills. & 0\% & 0\% & 11\% & 44\% & 44\%  \\
 \midrule 
Q3. Solving customized/personalized drag-and-drop programming questions was useful for my learning. & 0\% & 6\% & 11\% & 39\% & 44\%  \\
 \midrule 
Q4. Generating unlimited personalized drag-and-drop programming questions was useful for my learning. & 0\% & 0\% & 22\% & 28\% & 50\% \\
 \bottomrule
\end{tabular}
\end{table}

\begin{table}[htb]
\caption{Open-ended survey questions} 
\label{table:tool_openended_survey} 
\footnotesize
\begin{tabular} { c p{4.5cm}  c }
 \toprule
 \textbf{\emph{n}} & \textbf{Question} & \textbf{Optional?} \\
  \midrule
  18 & Q1. Why do you (or do not) find the customization of context in programming questions interesting/enjoyable?
  You can potentially describe any instances you have encountered with the tool to help support your answer. & No \\
 \midrule
  18 & Q2. Describe the usefulness of being able to select programming concepts in this tool.
  You can potentially describe any instances you have encountered with the tool to help support your answer. & No \\
  \midrule
  7 & Q3. Any overall thoughts or comments about the customization/personalization of programming questions can be included here. & Yes \\
  \midrule
  62 & Q4. Do you have any other feedback you would like to provide about PuzzleMaker? & Yes \\
 \bottomrule
\end{tabular}
\end{table}

\subsection{Likert Scale Survey Responses}
\noindent
A total of 18 students engaged with the Likert scale survey questions and their responses are summarized in Table \ref{table:tool_likert_survey}.


Students generally agreed that the ability to select their own context (Q1) was enjoyable. This supports the idea that personalizing Parsons problems was enjoyable, likely enhancing the learning experience of introductory programmers. Including various contexts contributes to positive feedback, suggesting that limiting contexts in a similar tool may not yield the same level of positive response, as it may fail to align well with the diverse range of student interests.

Additionally, they responded positively regarding the value of selecting programming concepts (Q2) for improving their programming skills. Moreover, students were very positive about the usefulness of solving personalized Parsons problems for their learning. Furthermore, they were very positive about generating an unlimited number of personalized Parsons problems being useful for their learning. 

The tool allows students to combine contexts and programming concepts in exercises, likely maintaining student enjoyment and exposing them to relevant learning materials. 
In response to our second research question (RQ2), the findings suggest that students generally responded positively to the tool, implying that they find contextualized programming exercises in the form of drag-and-drop (Parsons problems) interesting and useful for their learning.

\subsection{Open-ended Survey Responses}
The data from four open-ended questions (see Table \ref{table:tool_openended_survey}) were explored using thematic analysis. Relevant responses were extracted as codes and grouped into themes by the first author~\cite{braun_using_2006}. 

\subsubsection{Customisation of Contexts}
The question relevant to the inclusion of contexts (Q1) had four themes found within student responses (enjoyment, self-expression, value, and counterproductivity).  

The first theme, \textbf{enjoyability}, captures how context makes solving small programming exercises more pleasant than the typical experience of working with programming exercises devoid of context. Responses that illustrate this theme were \emph{“It makes the learning process more enjoyable and interesting if the context in which I am trying to solve the problem is something that interests me instead of some random generic problem.”} and \emph{“It's easier to work with something you're interested in”}. 

The second theme acknowledged that generating questions via the tool enabled \textbf{self-expression}. This theme describes students being given the creative freedom and control to create programming questions that align with their interests, with responses such as \emph{“I can come up with programming questions involving topics that I like and stuff I have as a hobby”}, \emph{“Because we can choose what our code is about.”}, and \emph{“because it gives us a sense of ‘oh we created this problem’”}. 

The third theme indicated that students believed that context was  \textbf{valuable} for learning. Responses associated with this theme were \emph{“It can be helpful to explain the context to understand the problem more and find what would be the best structure for the code to resolve the problem”} and \emph{“Was helpful to relate puzzles to different concepts”}. 

The final theme was more negative and encompassed the view that contexts could be \textbf{counterproductive}. Specifically, contexts may introduce added difficulty into the interpretability of problem statements, as witnessed by the comment \emph{“unnecessary context might be distracting from the question at hand”}.

\subsubsection{Customisation of Programming Concepts}
In the open-ended responses regarding the value of programming concepts (Q2), three themes emerged from student feedback. The first theme focuses on students' \textbf{learning opportunity}, for repeated practice of programming concepts. Practice questions may serve as a way to enhance understanding or serve as a general refresher for students, as indicated by the comments \emph{“It is useful because it enables targeted revision”}, \emph{“It lets us practice concepts that we may be struggling with, or also just help solidify anything that we have just learned.”}, and \emph{“it helps as I can focus on the areas that I feel less confident in and work through them at my own pace”}. 

The second theme was \textbf{exposure}. This theme highlights an opportunity for introductory programmers to encounter new concepts or ideas for solving problems they may not have seen or thought about before. Comments aligned with this theme included \emph{“it is nice to be able to see how a question should be answered if you are struggling with how to code something. examples are incredible at helping you see what you should do and where you can implement the same things in your code.”}, and \emph{“I can experience more problems and further my programming skills and resolve problems more efficiently, if i ever encounter the same problem to recall what I should be doing”}.

The final theme for this question was \textbf{limited}, describing some of the tool's inflexibility with applying code solutions. A common observation is that problems often have multiple solutions, as indicated by the comment \emph{“I didn't like their solution - I wanted to have the ability to provide my own solution.”}. Additionally, the tool integrated programming concepts into code solutions but failed to do something unique with them, as suggested in the following comment \emph{“I think this tool would be useful if I was struggling to understand a specific concept in programming and this tool would help in explaining the concept in an interesting way”}. 

\subsubsection{Personalisation of Parsons Problems}
There were relatively few responses to Q3, resulting in only two themes.  The first theme was \textbf{useful}, indicating how students saw personalized questions as a practical additional learning resource reinforcing their understanding of programming knowledge. Comments supporting this theme included \emph{“I thought it was great overall, a really good tool very helpful for learning and quite fun. Nice to use the drag and drop for understanding the structure of your code.”}, and \emph{“I would use that to better understand how I should structure my code and what I should write in my code to get the output that I want.”}.  In contrast to the \textbf{useful} theme,  the final theme was \textbf{undemanding},  describing the questions as lacking variety and not reflecting the difficulty of real-world programming problems. Responses reflecting this idea included \emph{“The questions seemed to be quite easy, maybe try make them harder.”}, and \emph{“Will it be possible to incorporate challenging questions that can be encountered in daily programming? Or show different ways to solve the problem.”}.  One student also remarked \emph{“Needs more complex questions exam style”}, highlighting a mismatch between the Parsons problems format and the typical code writing format used in exams.

\subsubsection{Other Tool Remarks}
Three themes were identified from student responses to the final optional open-ended question (Q4). The first theme was \textbf{tool issues}, which included the tool's appearance, functionality, and the limitations of basic Parsons problems compared to writing code from scratch. Comments reflecting this theme include \emph{“Upgrade the UI, maybe adding problems in other programming languages if possible”}, \emph{“I wish the site would behave more like coderunner where it can check the code that I wrote instead of the drag-and-drop format its in right now”}, \emph{“I think rather than drag and drop, it should make the user to actually write the code - similar to lab exercises.”}, and \emph{“I also feel like it's less useful as revision than trying to write the code yourself.”}. 

The second theme was \textbf{learning opportunity} and included comments such as \emph{“Personally I think this is a useful tool as some people like myself can't really just make up our own puzzles to practice on.”}, \emph{“It's helpful for visualizing and practicing code structure.”}, and \emph{“I found it somewhat helpful, it is a nice alternative to study”}.


The last theme was students finding \textbf{misleading} code solutions that do not match what they were explicitly taught in their introductory programming courses. Comments illustrating this theme were \emph{“Mine didn't seem to have the code written as best practice”}, and \emph{“Some code I had not used before so it was bit confusing”}.



\section{Conclusion}
We developed an educational tool, PuzzleMakerPy, that leverages a large language model for automating the generation of Parsons problems with customizable programming concepts and thematic contexts.  We deployed PuzzleMakerPy in a large introductory programming course, finding that 
students viewed the ability to customize Parsons problems with contexts as enjoyable, and reported the ability to select programming concepts as useful for their learning. We found that students were enthusiastic about being able to create programming questions that aligned with their interests.  They engaged with a wide variety of available contexts, some even typing their own, and few requested Parsons problems without context.  Students also made use of the programming concepts provided to them in a small predefined list but did not choose to type their own. 
Overall, our findings highlight the strengths and benefits of PuzzleMakerPy, particularly how the contextualization of Parsons Problems can be enjoyable for introductory programmers while maintaining learning benefits by allowing customization of programming topics.  Future work should continue to explore the potential for automatically generating personalized learning resources for students, especially by leveraging large language models.


\section{Acknowledgements}


This study was conducted with the ethics approval of \emph{University of Auckland Human Participants Ethics Committee (UAHPEC)}, with reference \emph{UAHPEC25279}.


\balance


\bibliographystyle{ACM-Reference-Format}
\bibliography{references}

\end{document}